\documentclass[12pt]{article}
\usepackage[utf8]{inputenc}
\usepackage{geometry}
\usepackage{graphicx}
\usepackage{amsmath}
\usepackage{amssymb}

\let \oldvec \vec
\renewcommand{\vec}[1]{\oldvec{\mathbf{{#1}}}}

\newcommand{\uv}[1]{\, \hat{\mathbf{#1}}}

\begin{document}

\title{A numerical study of approximate metrics with quadrupole moment}
\author{Guillermo Andre Oliva--Mercado\thanks{Escuela de Física, Universidad de Costa Rica. {\ttfamily guillermo.olivamercado@ucr.ac.cr} and  {\ttfamily gandreoliva@gmail.com}}, Francisco Frutos--Alfaro \thanks{Centro de Investigaciones Espaciales y Escuela de Física, Universidad de Costa Rica. {\ttfamily francisco.frutos@ucr.ac.cr}},\\
 Javier Bonatti--González\thanks{Centro de Investigación en Ciencias Atómicas, Nucleares y Moleculares; y Escuela de Física. {\ttfamily javier.bonatti@ucr.ac.cr}}}
\date{25 February 2016}
\maketitle

\textbf{Abstract}

{\small Recently, spacetimes described by metrics with three parameters (mass, rotation and small quadrupole moment) was found, and in this paper, null geodesics for these metrics are calculated and visualized. Light scattering, as well as the role that the quadrupole moment plays in deforming these kinds of spacetimes are studied. This is a new application of a symbolic--numeric program that we previously used to study the Bonnor metric.}

\textbf{Resumen}

{\small Recientemente se han encontrado espacio-tiempos descritos por métricas con tres parámetros (masa, rotación y momento de cuadrupolo pequeño), y en este artículo, se calculan y visualizan las geodésicas nulas para estas métricas. Se estudia la dispersión de la luz, así como el papel que juega el momento de cuadrupolo en estos tipos de espacio-tiempos. Esta es una nueva aplicación del programa simbólico--numérico que previamente usamos para estudiar la métrica de Bonnor.}

\vspace{0.5cm}


\textbf{Palabras clave:} Relatividad General, Geodésicas, Geometría Diferencial, Software Libre.

\section{Introduction}
In General Relativity, geodesics represent the trajectories of massive and massless particles (light). They are calculated by means of solving the geodesic equations, which is a set of second order ordinary differential equations for $x_\mu(\lambda)$, where $\lambda$ is an affine parameter and $\mu=0,1,2,3$. The geodesic equations contain the Christoffel symbols, which contain first order partial derivatives of the metric.

Null geodesic visualization can be used to study the properties of spacetime \cite{hagihara30}, as well as modeling high resolution gravitational lenses and radiation near compact objects \cite{cunningham75, luminet79, yang-14}. In a previous effort, we devised a program that, in principle, takes any given metric and evaluates null geodesics for given initial conditions; we then applied it to the Schwarzschild and Bonnor metrics \cite{oliva15}.

Previous works on null geodesics used manual analytical calculations of the geodesic equations and then numerical evaluation \cite{cunningham75,luminet79}; others even integrate once analytically and the remaining first-order differential equation is numerically evaluated, producing high-speed codes \cite{dexter-agol10, yang-14}, even GPU-accelerated \cite{muller11}. Entirely numerical metrics using the 3+1 formalism of General Relativity can also be studied using the code in \cite{vincent-11}. The resulting code of these works, with the exception of the last one, can be applied to only a limited subset of metrics, including the Schwarzschild and Kerr metrics.

The Kerr metric, however, has difficulties to be matched to a realistic interior solution and might not accurately represent any external field of a real astrophysical object \cite{dymnikova15,frutos15}; with this in mind, an approximate metric was obtained by Frutos et al. which includes a quadrupole term of first order \cite{frutos15}. Another approximate metric that contains quadrupole terms of second order was also proposed\cite{frutos15a} and this is the one that we analize with greater detail here.

This paper is structured as follows: first, we present the metrics being considered in the results; then, we discuss the program, method and initial conditions; later we present the results and limit the behavior of geodesics to certain regions depending on the impact parameter, ending with a comparison between the first and second order expansions for the metric.

\section{Metrics}
The metric with second order quadrupole terms is \cite{frutos15a}

\[ ds^2 = g_{tt}dt^2 + 2g_{t\phi} dtd\phi + g_{rr} dr^2 + g_{\theta\theta} d\theta^2 + g_{\phi\phi}d\phi^2 \]
\begin{align}
\begin{split}
g_{tt} &= \frac{e^{-2\psi}}{\rho^2} [a^2\sin^2\theta - \Delta]\\
g_{t\phi} &= -\frac{2Jr}{\rho^2}\sin^2\theta\\
g_{rr} &= \rho^2 \frac{e^{2\chi}}{\Delta}\\
g_{\theta\theta} &= \rho^2 e^{2\chi}\\
g_{\phi\phi} &= \frac{e^{2\psi}}{\rho^2} [ (r^2+a^2)^2 - a^2\Delta \sin^2\theta ]\sin^2\theta
\end{split}\label{generalmetric}
\end{align}
where
\begin{align}\begin{split}
\psi &= \frac{q}{r^3}P_2 + 3 \frac{Mq}{r^4}P_2 \\
 \chi &= \frac{qP_2}{r^3} + \frac{Mq}{r^4}\left( -\frac{1}{3} + \frac{5}{3}P_2 + \frac{5}{3}P_2^2 \right) + \frac{q^2}{r^6}\left(\frac{2}{9}-\frac{2}{3}P_2 -\frac{7}{3}P_2^2 + \frac{25}{9}P_2^3 \right)
 \end{split}\label{frutosexpanded}
 \end{align}
are functions obtained by expansion, $P_2 = (3\cos^2\theta - 1)/2$, and
\[ \Delta =  r^2 - 2Mr + a^2 \]
\[ \rho^2 = r^2 + a^2\cos^2\theta \]
as in the Kerr metric. This metric has a validity up to order $O(aq^2,a^2q,Maq,Mq^2,M^2q,q^3)$, it has Kerr as limit, and may be matched with an interior solution.

The parameters of the metric are $M$, the mass; $a$, the angular momentum parameter (the angular momentum is $J=Ma$; $a<M$) and $q$, the quadrupole moment parameter, which controls the deformation of the object. We use geometrical units ($G=c=1$).

In a similar way, the metric with a linear quadrupole term has the same definitions as the one above but (see \cite{frutos15})
\begin{equation} \psi = \chi = \frac{2qM^3}{15}\frac{P_2}{r^3} \label{first-order}\end{equation}

\section{Program and numerical analysis}

As discussed in \cite{oliva15}, our program uses a symbolic--numeric algorithm written in Sage and Python; it first takes the given metric and computes the geodesic equations, writing a Python script that allows the numeric solution using the Runge-Kutta method of 4th order. A module specifies and varies the initial conditions, ensures that these conditions produce null geodesics, and makes the necessary coordinate transformations to represent the results in $\mathbb{R}^3$. There is another module for visualization using Gnuplot and Vpython, and another module for statistical analysis.

\subsection{Impact parameter vector and initial conditions}
\begin{figure}
\begin{center}
a) \includegraphics[width=6cm]{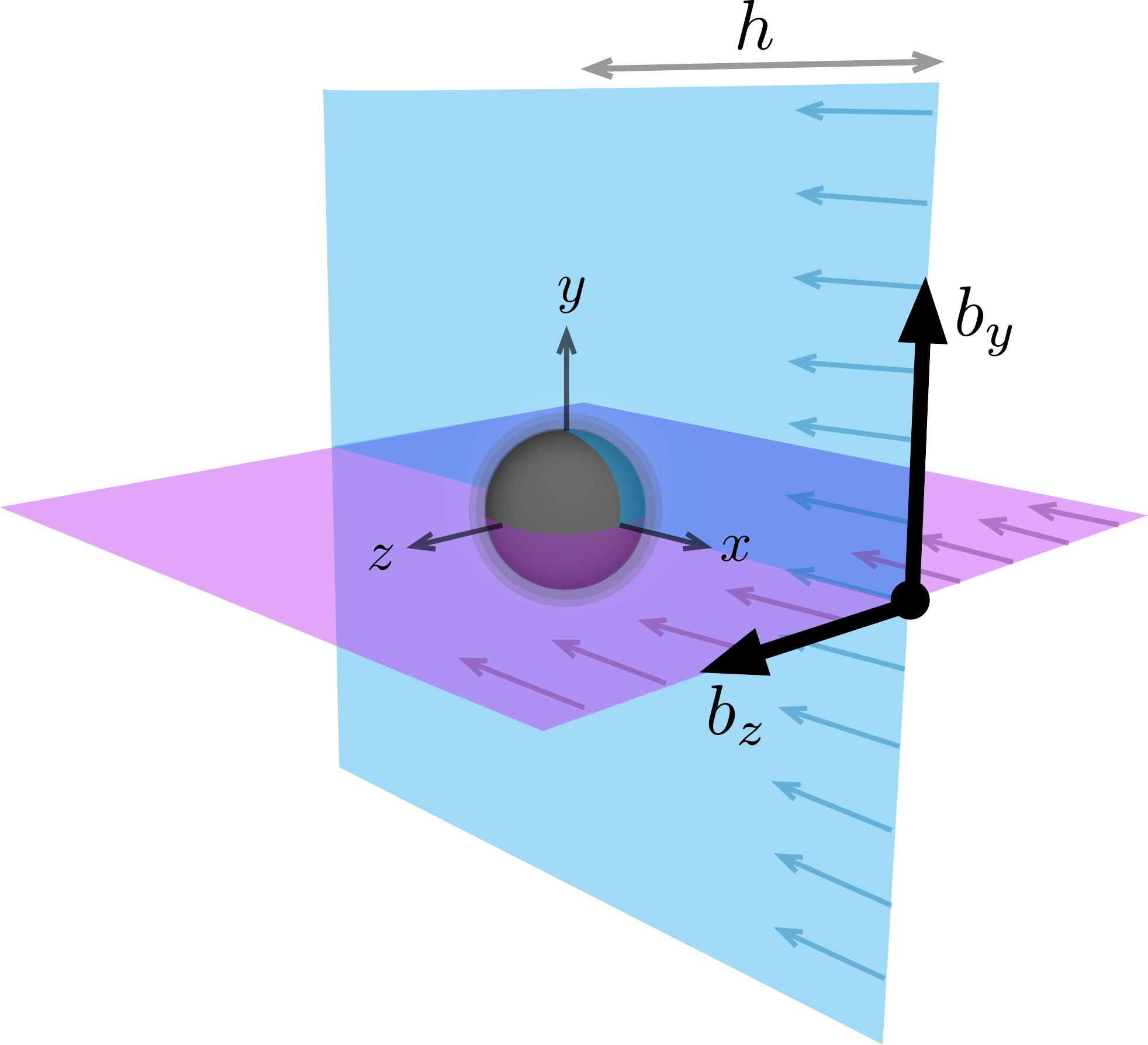}
b) \includegraphics[width=6cm]{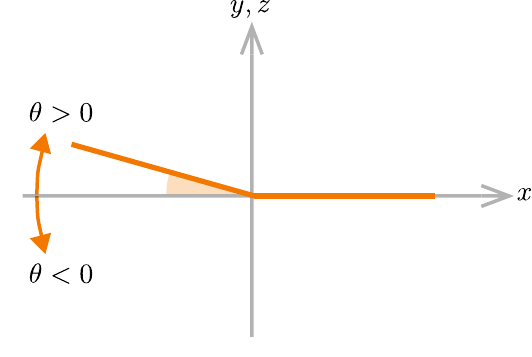}
\caption{a) Definition of the impact parameter vector. b) Definition of the scattered angle.}
\label{diagrama}
\end{center}
\end{figure}
Scattering of light by an object with deformation and rotation is non-symetrical. Therefore, we define an impact parameter vector $\vec b$ as in Fig. \ref{diagrama}a, in order to specify the position of the incident beam. The geodesics were all released in the $-\uv x$ direction from a distance $h=10$. In the Fig. \ref{diagrama}b we show our definition of the scattered angle $\theta$, which is measured in the projection to the $xy$- or $xz$-planes, depending on the case.

In this paper, we give values with an uncertainty that represents the interval at which the values of the components of $\vec b$ were varied. It is important to note that, for clarity, the graphics we show contain less geodesics than the ones used to obtain the results. Also, a dot or sphere marks the position of the compact object; the angular momentum goes in the $\uv z$ direction, and the Schwarzchild radius is $2$.

\section{Results}

\subsection{Variation with the impact parameter}
\begin{figure}
\begin{center}
a)\includegraphics[width=6cm]{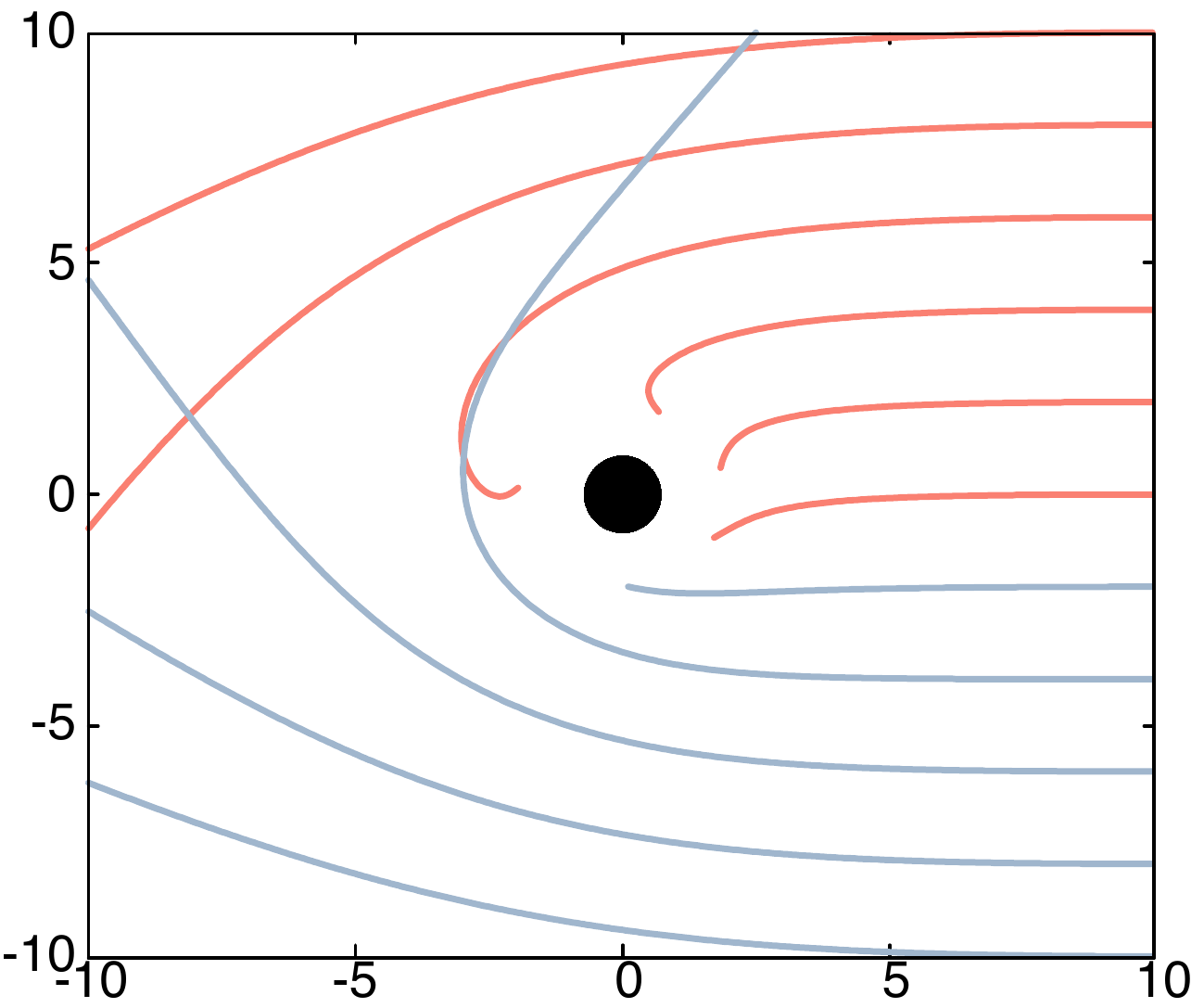}
b)\includegraphics[width=8cm]{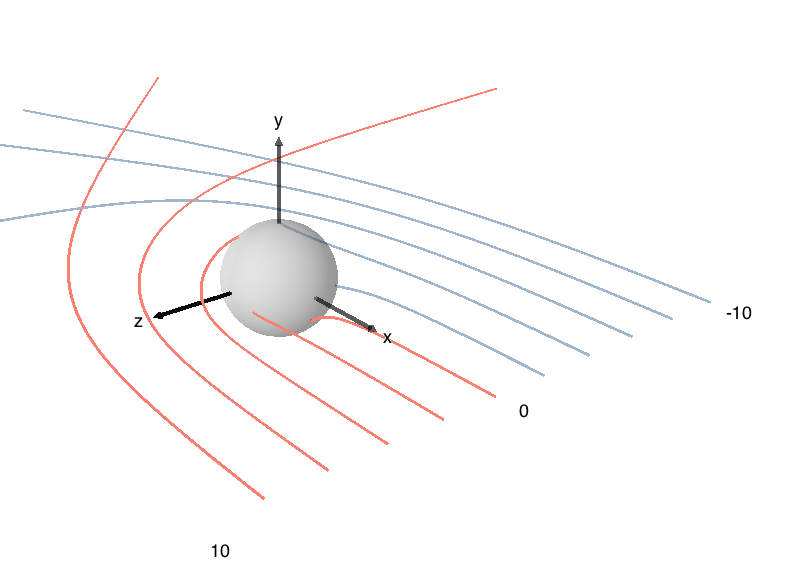}
\caption{Variation of the impact parameter in: a) the $xy$ plane; b) the $xz$ plane.}
\label{comp-b}
\end{center}
\end{figure}

We have identified several intervals of the components of $\vec b$ that describe the behavior of geodesics, represented trough the scattered angle $\theta$. For scattering in the $xy$-plane (Fig. \ref{comp-b}a), geodesics stay in the plane. Values of $b_y$ between $0$ and $6.4\pm 0.1$ produce a capture by the compact object. For $6.4\pm 0.1 < b_y < 7.1 \pm 0.1$, geodesics escape with $\theta < -90^\circ$, that is, some geodesics get reflected and cause a ``gravitational mirror'' effect. With values of $b_y$ greater than $7.1\pm 0.1$, geodesics are scattered with $\theta > -90^\circ$, and this angle keeps reducing with distance; for $b_y = 13.5\pm 0.5$ it is $\theta \approx -18^\circ$.

The situation in the ecuatorial ($xy$-) plane is asymetrical, because of the frame dragging produced by the rotation of the object. Because of this, the negative part of the Fig. \ref{comp-b}a shows different results. In the interval $-4.8 \pm 0.1<b_y<0$, the geodesics get scattered or captured with an angle $\theta > 90^\circ$. For values of $b_y$ lower than $-4.8\pm 0.1$, the angle is less than $90^\circ$, and keeps reducing such that for $b_y = -12.0\pm 0.5$, $\theta \approx 16^\circ$.

A similar analysis was also done for geodesics in the $xz$-plane. However, these geodesics do not stay in the plane, swirling around the event horizon and/or escaping with a deviation instead. The results in this case were quite simetric: for a $0<|b_z|<6.0\pm 0.1$, geodesics get captured or escape with $|\theta|>90^\circ$. For values greater than $6.0 \pm 0.1$, geodesics scatter with $|\theta|<90^\circ$ and the angle keeps reducing such that for $|b_z|=12.5\pm 0.5$, $|\theta|\approx 15^\circ$.

It is important to recall that $h=10$, and with a higher value of $h$, the results will be more accurate since the rays have to come parallel from infinity.

\subsection{Scalability of the results}
\begin{figure}
\begin{center}
a)\includegraphics[width=6cm]{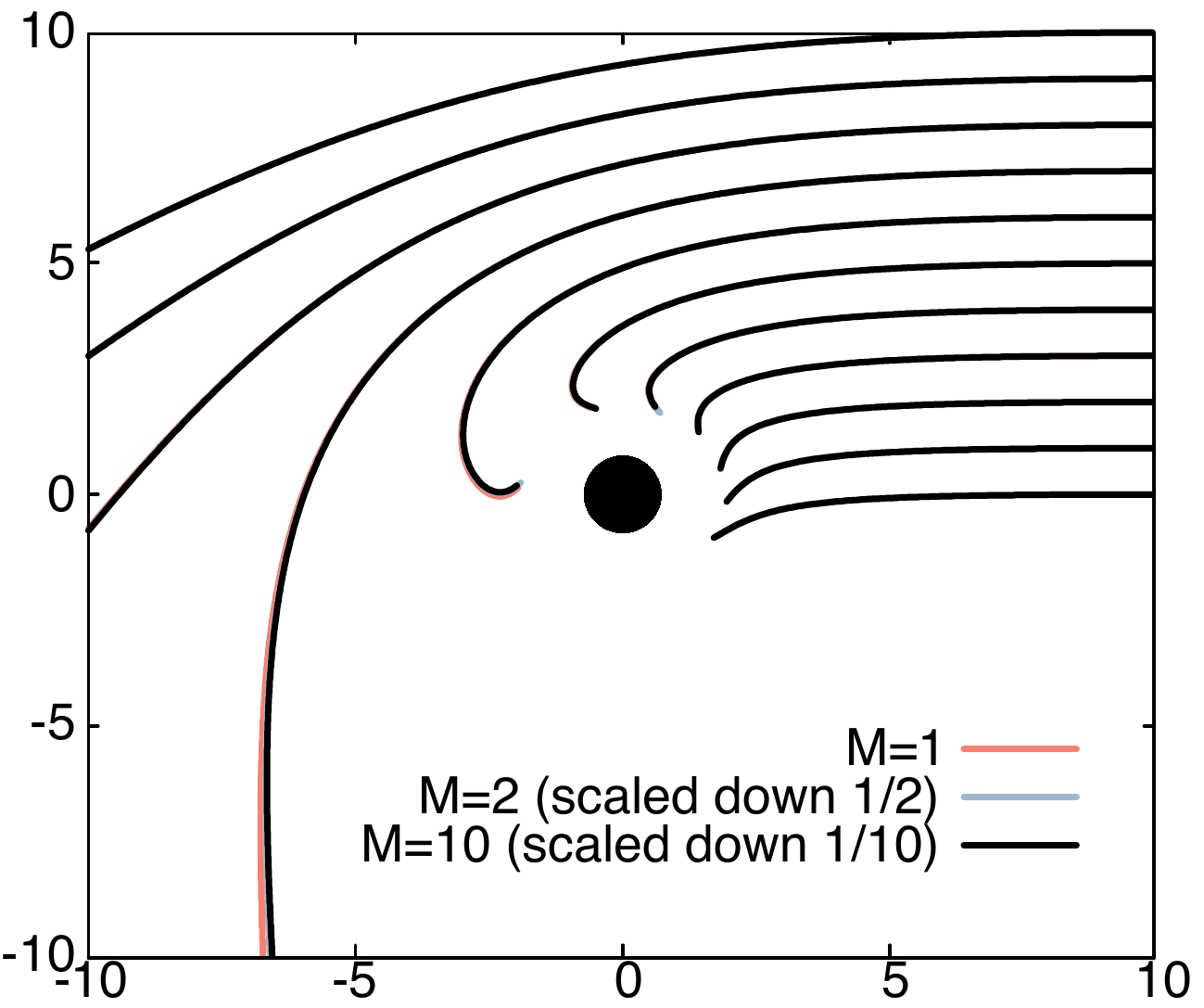}
b)\includegraphics[width=6cm]{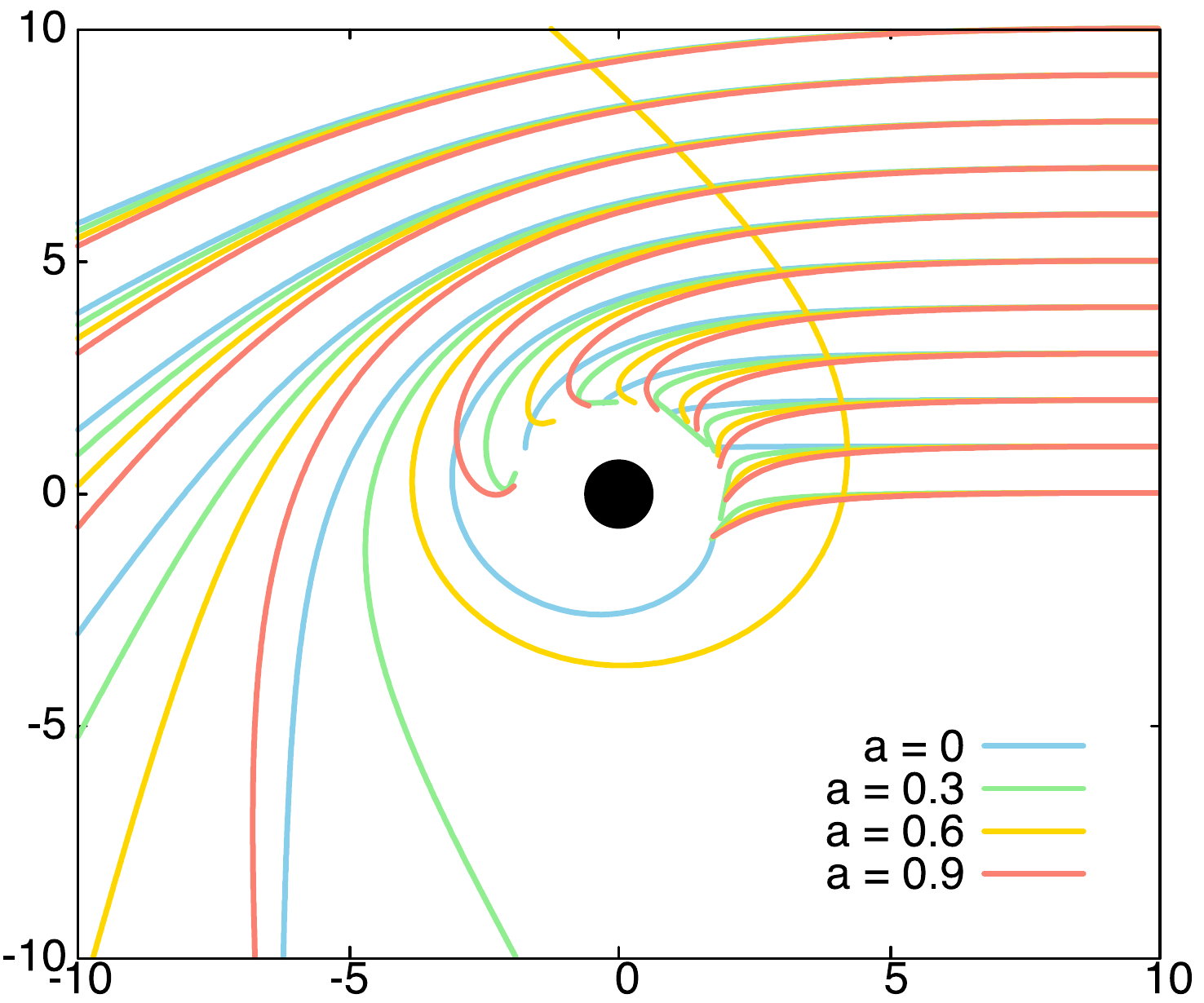}
\caption{a) Scalability of the results b) Variation with $a$. Both plots represent the $xy$-plane.}
\label{comp-m}
\end{center}
\end{figure}
Our use of geometrical/Planckian units requires a further exploration of the physical applicability of our results. To this end, we tested masses $M=1,2,10$, but also varied proportionally all parameters of the metric ($q=0.1M$, $a=0.9M$), and of the simulation. Then, to compare, we plotted the results but scaled down by exactly the same proportion. The results are in Fig. \ref{comp-m}a, and show no appreciable difference, as expected. This means that our results are indeed applicable to astrophysical objects.

\subsection{Variation with the angular momentum parameter}
The metric (\ref{generalmetric}) with (\ref{frutosexpanded}) reduces to the Schwarzschild metric when $a=q=0$. However, we wanted to explore the behavior of the parameter $a$ with $M=1$ and $q=0.9$. The results are in Fig. \ref{comp-m}b: they show the effect of reducing the angular momentum of the compact object. The well-known Kerr-like frame dragging is progresively reduced, until a Schwarzschild-like behavior is achieved. It is worth noting that with increasing $a$, geodesics tend to swirl and escape in angles $\theta \sim -180^\circ$.

\subsection{Variation with the quadrupole moment parameter}
\begin{figure}
\begin{center}
a)\includegraphics[width=6cm]{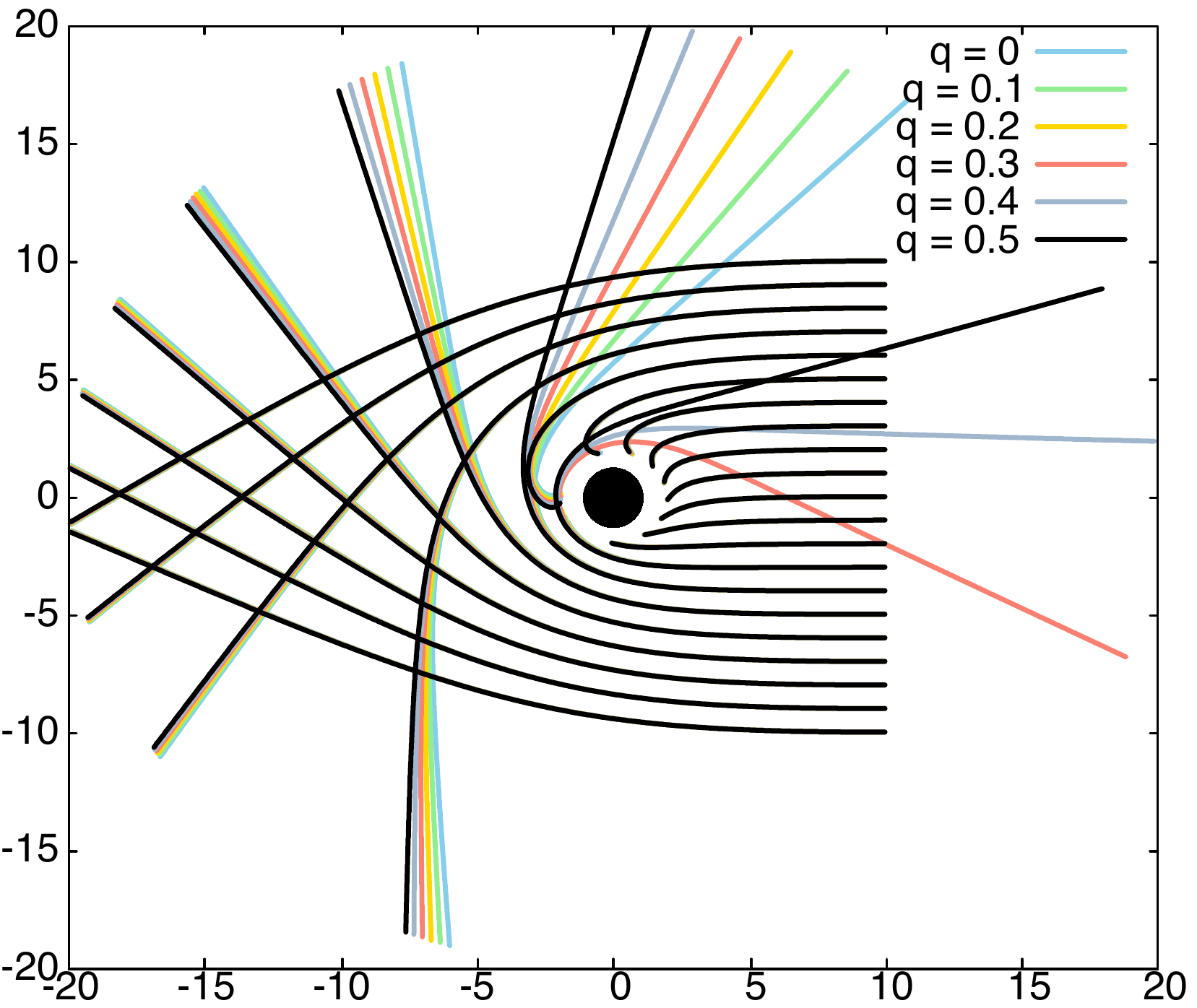}
b)\includegraphics[width=6cm]{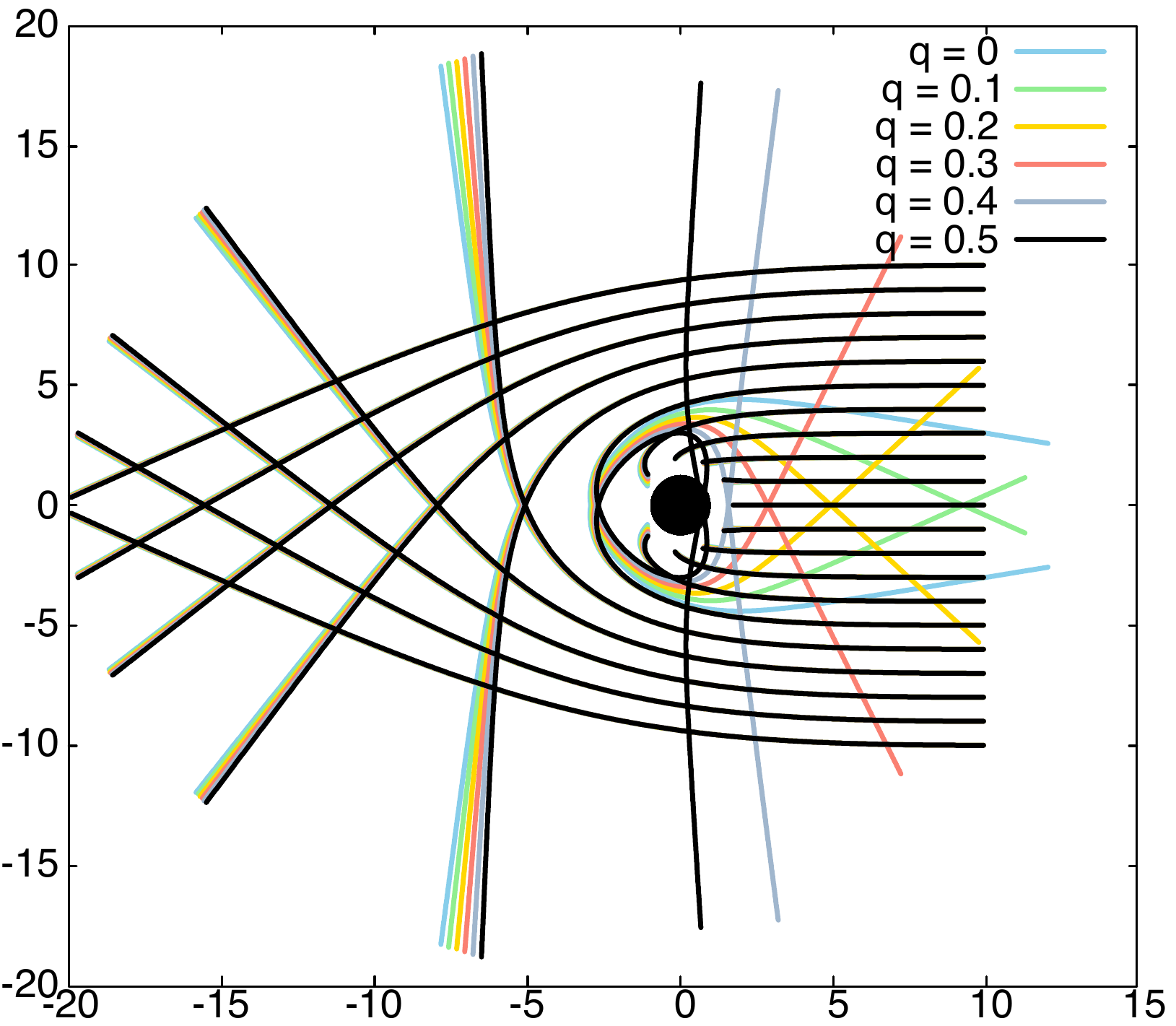}
c)\includegraphics[width=6cm]{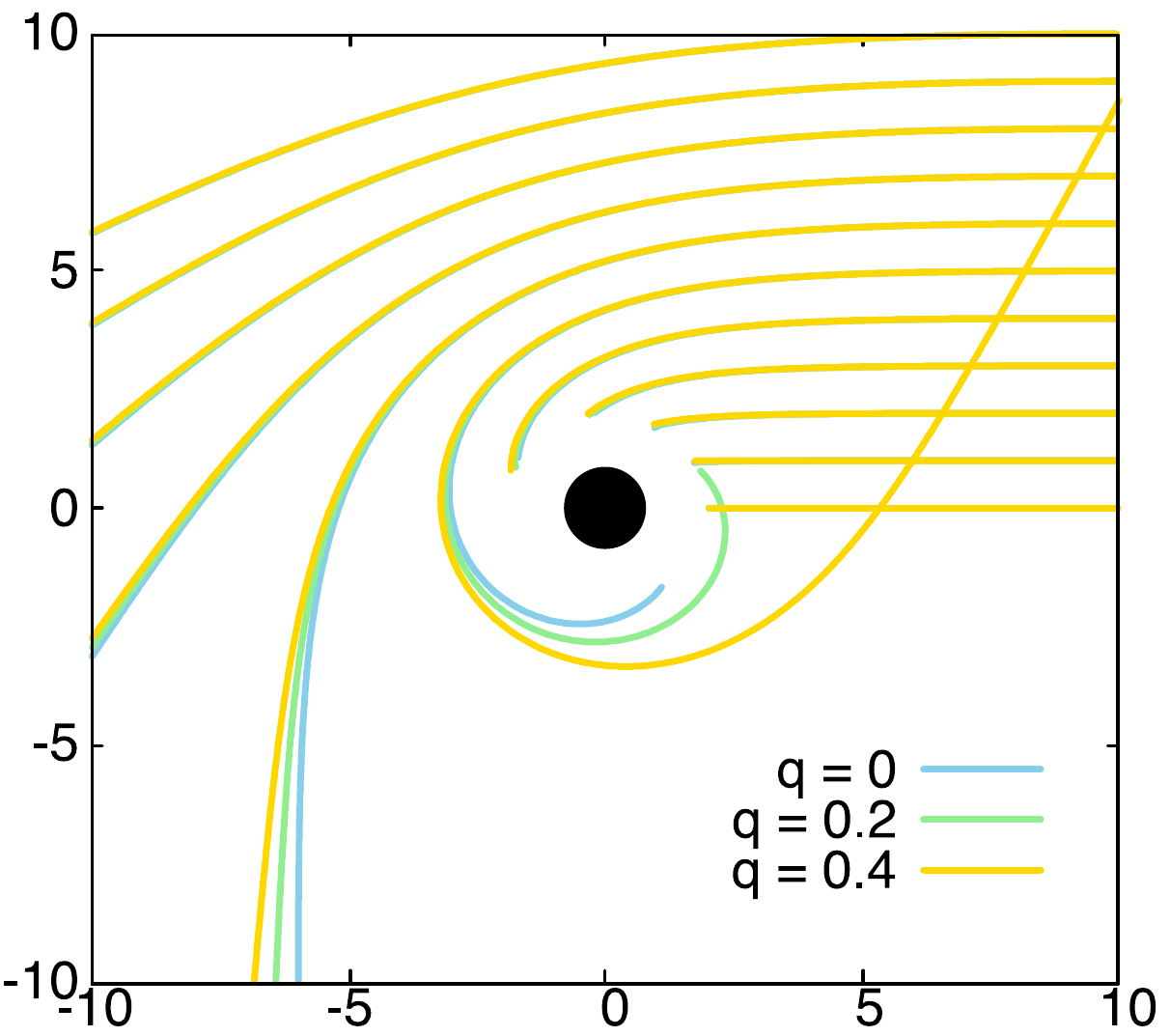}
\caption{Variation of the quadrupole moment parameter in: a) the $xy$-plane; b) the $xz$-plane; c) the $xy$-plane with $a=0$. These plots are for the metric with a second order expansion in $q$.}
\label{comp-q}
\end{center}
\end{figure}

Now we would like to discuss the effect of the parameter $q$ on the metric. We already mentioned that with $q=0$, we recover the Kerr metric. In the figures \ref{comp-q}a and \ref{comp-q}b, we see that geodesics that scatter with $|\theta| \gtrsim 90^\circ$ have the most difference with different values of $q$. Also, scattering in the $xz$-plane has the greatest difference with $q$ for the rays that get reflected with $\theta \sim 180^\circ$. Theoretically, this difference could be used to measure the deformation of an object using the reflected rays. Also, it allows us to infer that gravitational lenses might not be suitable to measure such small deformations.

Finally, we would like to compare the ecuatorial geodesics without rotation ($a=0$) to the Schwarzschild metric, that is, a non-rotating deformed object. Fig. \ref{comp-q}c shows the results. Again, the differences are most noticeable for rays with $|\theta| > 90^\circ$. It is important to mention that we estimated the numerical error produced by the method and checked these results against it; at the end of the calculations, the deviation effect of $q$ is at least two orders of magnitude larger than the numerical error, and therefore, the geodesics are indeed distinct.

\subsection{Comparison with the first order metric}
\begin{figure}
\begin{center}
a)\includegraphics[width=6cm]{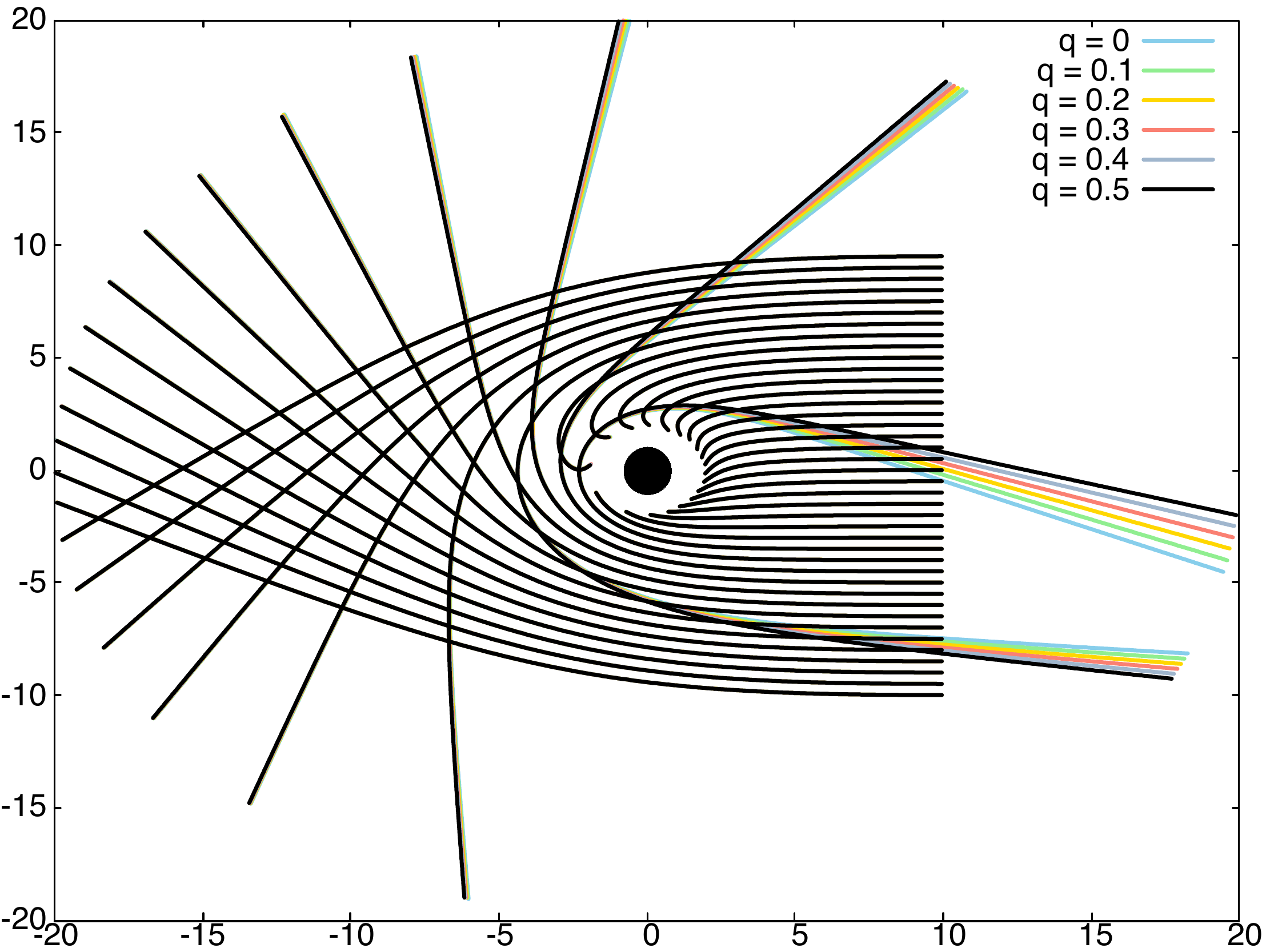}
b)\includegraphics[width=6cm]{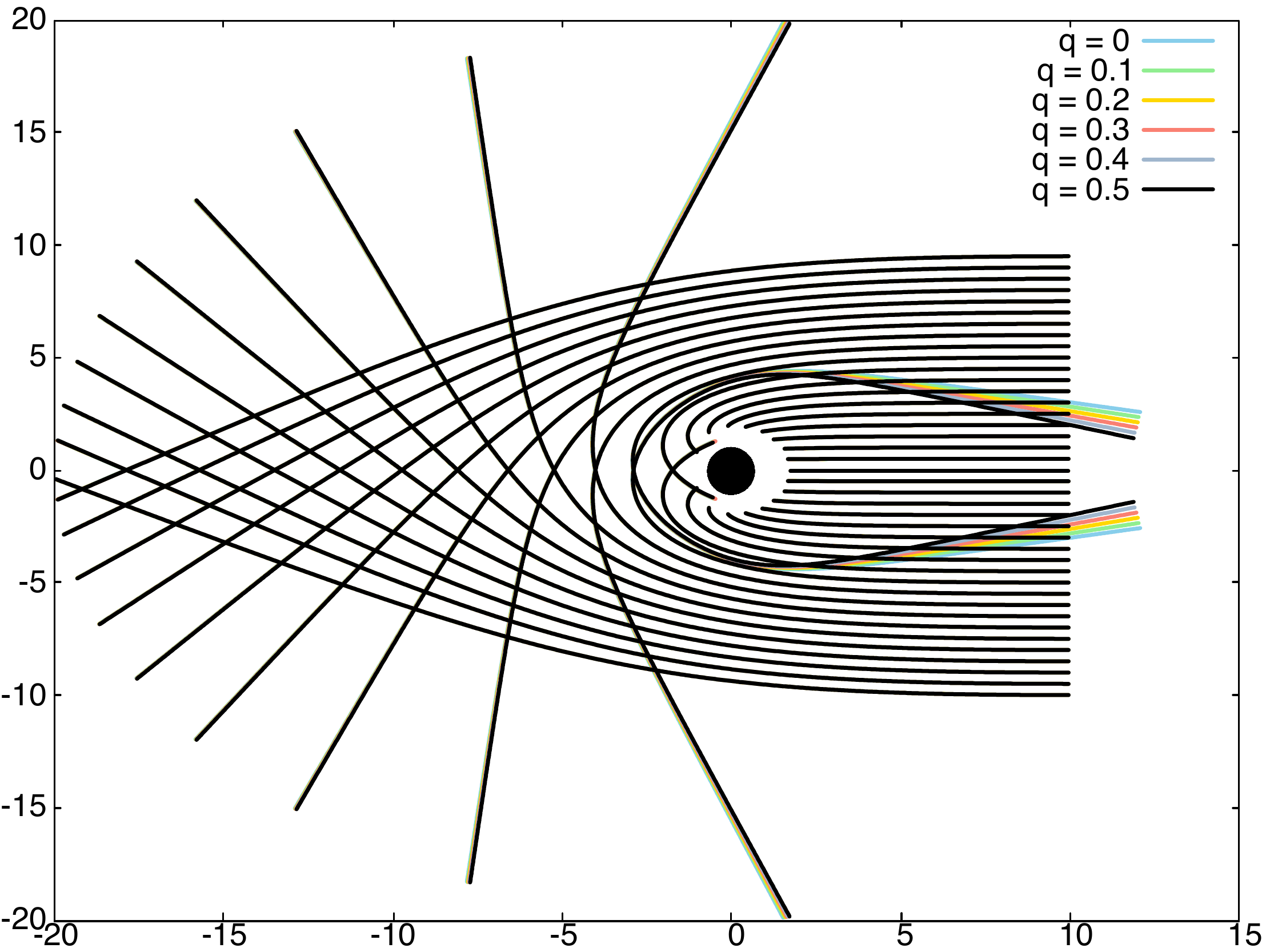}
\caption{Variation of the quadrupole moment parameter in: a) the $xy$-plane; b) the $xz$-plane. These results are for the metric with a first order expansion in $q$.}
\label{comp-q-fo}
\end{center}
\end{figure}

In the Fig. \ref{comp-q-fo}  we study the variation with $q$ for the metric (\ref{generalmetric}) with (\ref{first-order}). We see that the differences between values of $q$ are weak in both cases. Comparing with the Fig. \ref{comp-q}, the differences are much weaker ---specially for the case of the $xz$-plane--- with the first order expansion. Again, the reflected and strongly deviated rays are the key to possible measurements, although detecting differences in this case is much more difficult.

\section{Conclusions}
We presented a visualization of two Kerr-like metrics by studying the scattering of light around them. This constituted another application case of our program. We found that the effects on the null geodesics of adding a first and second order expansion of $q$ in the metric are small, but theoretically measurable using the reflected rays, although much more difficult using gravitational lenses. We also identified several intervals of the components of $\vec b$ and their corresponding intervals of the scattered angle.

\vspace{2cm}

\textbf{Acknowledgements}. The authors are grateful to the following libre software projects:

Sage Mathematics Software (Version 6.10), The Sage Developers, 2015,\newline {\ttfamily http://www.sagemath.org}.

Ginac, 2015 {http://www.ginac.de}. Included in Sage v6.10.

Vpython (Version 6.11), 2015. {\ttfamily http://vpython.org} 

Gnuplot (Version 5.0), 2015, {\ttfamily http://www.gnuplot.info}

\end{document}